\documentclass[fleqn,twoside]{article}
\usepackage{espcrc2}
\usepackage{epsf}

\usepackage{graphicx}
\usepackage[figuresright]{rotating}

\newcommand{\AmS}{{\protect\the\textfont2
  A\kern-.1667em\lower.5ex\hbox{M}\kern-.125emS}}

\def\la{\langle}
\def\ra{\rangle}

\def\beq{\begin{equation}}
\def\eeq{\end{equation}}
\def\bea{\begin{eqnarray}}
\def\eea{\end{eqnarray}}

\def\op{{\mathcal{O}}}

\def\chsu3{${\mathrm{SU}}(3)_{\mathrm{L}}\times {\mathrm{SU}}(3)_{\mathrm{R}}$}
\def\lsu3{${\mathrm{SU}}(3)_{\mathrm{L}}$}
\def\rsu3{${\mathrm{SU}}(3)_{\mathrm{R}}$}

\title{$K^{+}\rightarrow\pi^{+}\pi^{0}$ decays at next-to-leading order
in the chiral expansion on finite volumes \thanks{SHEP 02-18.}}

\author{
C.-J.D. Lin$^{a}$\thanks{Presenter at the conference.},
G. Martinelli$^{b}$,
E. Pallante$^{c}$,
C.T. Sachrajda$^{a,d}$
and
G. Villadoro$^{b}$\\
\vspace{0.3cm}
$^{a}$ Dept. of Physics and Astronomy, Univ. of Southampton,
       Southampton SO17 1BJ, England\\
$^{b}$ Dip. di Fisica, Universit\'{a} di Roma ``La Sapienza'',
       Piazzale A. Moro 2, I-00185 Roma, Italy\\
$^{c}$ SISSA, Via Beirut 2-4, 34013, Trieste, Italy\\
$^{d}$ Theory Division, CERN, CH-1211 Geneva 23, Switzerland
}
       
\begin{document}

\begin{abstract}
We present the ingredients for determining 
$K^{+}\rightarrow\pi^{+}\pi^{0}$ matrix elements via the combination
of lattice QCD and chiral perturbation theory ($\chi$PT).  By simulating
these matrix elements at unphysical kinematics, it is possible to 
determine all the low-energy constants (LECs) for constructing 
the physical $K^{+}\rightarrow\pi^{+}\pi^{0}$ amplitudes at next-to-leading
order (NLO) in the chiral expansion.
In this work, the one-loop chiral corrections are calculated for
arbitrary meson four-momenta, in both $\chi$PT 
and quenched $\chi$PT (q$\chi$PT), and the
finite-volume effects are studied.
\vspace{1pc}
\end{abstract}

\maketitle

\section{INTRODUCTION}
The need for a 
high-precision prediction for $K\rightarrow\pi\pi$ amplitudes is
underlined by the recent experimental measurement of 
Re($\epsilon^{\prime}/\epsilon$) and the long-standing puzzle, the 
$\Delta I = 1/2$ rule.  Although the finite-volume (FV)
techniques developed in Refs. \cite{Lellouch:2000pv,Lin:2001ek,Lin:2001fi}
can ultimately enable an accurate calculation of $K\rightarrow\pi\pi$ decay
rates, the most practical approach to the numerical
calculation of these decay rates remains the combination of lattice QCD and
(quenched and partially quenched) $\chi$PT.  
Apart from a calculation for the CP-conserving, $\Delta I = 3/2$, 
$K\rightarrow\pi\pi$ decay in Ref. \cite{Aoki:1997ev}, 
all the numerical studies 
hitherto follow a strategy proposed
in Ref. \cite{Bernard:wf}, which only allows the determination of these
amplitudes at leading-order (LO) in the chiral expansion\footnote{In Ref. 
\cite{Aoki:1997ev}, the decay amplitude is also obtained at the 
precision of leading-order in the chiral expansion.}.  Because of the 
large kaon mass and the presence of 
final state interactions, non-LO corrections in
this expansion are significant.
In a recent work
\cite{Lin:2002nq,Boucaud:2001mg}, we have proposed to perform lattice
simulations at unphysical kinematics over a range of meson masses and momenta,
from which we can determine all the necessary LECs for constructing the
physical matrix element $\la\pi^{+}\pi^{0}|\op^{\Delta S = 1}|K^{+}\ra$ at 
NLO in the chiral expansion.  
We have suggested a specific
unphysical kinematics\footnote{Another choice 
is considered in Ref. \cite{Laiho:2002jq}.}, the SPQR kinematics
as explained in detail in Refs. \cite{Lin:2002nq,Boucaud:2001mg}, which
enables such a procedure, and have studied the 
following $\Delta S = 1$ operators
($\alpha$, $\beta$ are colour indices)
\vspace{-0.1cm}
\begin{eqnarray}
 Q_{4} &=&  
  (\bar{s}_{\alpha}d_{\alpha})_{L}(\bar{u}_{\beta}u_{\beta}
  - \bar{d}_{\beta}d_{\beta})_{L}\nonumber\\
 & &+ 
  (\bar{s}_{\alpha}u_{\alpha})_{L}(\bar{u}_{\beta}d_{\beta})_{L} ,\nonumber\\
\label{eq:OpDef}
 Q_{7} &=& \frac{3}{2}(\bar{s}_{\alpha}d_{\alpha})_{L}
      \sum_{q=u,d,s}e_{q}(\bar{q}_{\beta}q_{\beta})_{R} , \\
 Q_{8} &=& \frac{3}{2}(\bar{s}_{\alpha}d_{\beta})_{L}
      \sum_{q=u,d,s}e_{q}(\bar{q}_{\beta}q_{\alpha})_{R} ,\nonumber
\end{eqnarray}
where $e_{q}$ is the electric
charge of $q$ and $(\bar{\psi}_{1}\psi_{2})_{L,R}$ means
$\bar{\psi}_{1}\gamma_{\mu}(1\mp\gamma_{5})\psi_{2}$.

\section{FINITE-VOLUME EFFECTS}
In Ref. \cite{Lin:2002nq}, we investigate the FV corrections,
power-like in $1/L$, which arise from replacing sums by integrals in the 
one-loop calculation that involves the diagrams in Fig. 1
\footnote{
Such a calculation for 
$\la\pi^{+}\pi^{0}|Q_{4}|K^{+}\ra$
at two particular kinematics, $M_{K}=M_{\pi}$ and $M_{K}=2 M_{\pi}$
with all mesons at rest,
have been performed in Refs. \cite{Golterman:1997wb,Golterman:1998af}.
}.
It can be shown that a diagram which does not have an imaginary
part in Minkowski space will only have FV corrections
exponential in $L$, therefore only diagram (c) 
contributes to the $1/L^{n}$ corrections.  Because the two-pion final
state has $I=2$, this diagram only contains four-quark 
intermediate states and there are no disconnected
quark-loops in the quark-flow picture.
For the same reason, it does not receive contributions 
from the $\eta^{\prime}$ propagator.  Hence the $1/L^{n}$ effects
are identical in $\chi$PT and quenched $\chi$PT (q$\chi$PT) 
for $K^{+}\rightarrow\pi^{+}\pi^{0}$ 
at this order, and only at this order.  This is not true for
$\Delta I = 1/2$ decay amplitudes \cite{Villadoro:latt02,Golterman:1999hv}.
\vspace{-1cm}
\begin{center}
\begin{figure}[hbt]
\label{fig:diagrams}
\epsfxsize=5cm
\epsffile{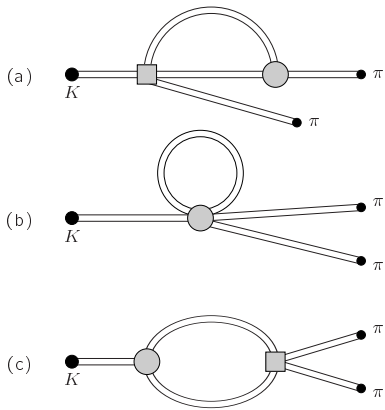}
\vspace{0.0cm}
\caption{{\sl One-loop diagrams for $K^{+}\rightarrow\pi^{+}\pi^{0}$
amplitudes.  The grey circles (squares) are weak (strong) vertices.  
The diagrams for wavefunction renormalisation
are not shown here.}}
\end{figure}
\end{center}
\vspace{-1.2cm}

\smallskip

We have found that in the center-of-mass frame, for all the 
$K^{+}\rightarrow\pi^{+}\pi^{0}$ amplitudes, these one-loop 
$1/L^{n}$ corrections are independent of
the weak operators and can be removed by a universal factor
derived first by Lellouch and L\"{u}scher
in Ref. \cite{Lellouch:2000pv}.  This
modifies the conclusion of Refs.
\cite{Golterman:1997wb,Golterman:1998af}, in which
a FV effect
resulting in the shift of the two-pion total energy in
the argument of the tree-level amplitudes is 
interpreted as a genuine $1/L^{n}$ correction to the matrix
elements, and therefore the FV effects in 
$\la\pi^{+}\pi^{0}|Q_{4}|K^{+}\ra$ are found to depend on $M_{K}$.

\smallskip

We are currently investigating the FV effects of 
these amplitudes in a moving frame. The
Lellouch-L\"{u}scher factor has not yet been derived for this, 
while the modification of L\"{u}scher's quantisation condition 
\cite{Luscher:1986pf,Luscher:1990ux} relating the infinite-volume $\pi\pi$
scattering phase to the FV two-pion energy spectrum,
due to the moving frame was obtained in Ref.
\cite{Rummukainen:1995vs}.  As a by-product of our work,
we verify that the energy shift 
obtained in one-loop perturbation theory in a moving frame agrees with
the expansion of the quantization condition in 
Ref. \cite{Rummukainen:1995vs} to the same order.

\section{ONE-LOOP CHIRAL CORRECTIONS IN INFINITE VOLUME}
We evaluate the one-loop correction by using
dimensional regularisation and 
subtracting ${\mathrm{log}}(4\pi) - \gamma_{E} + 1 + 2/(4-d)$.
The lowest-order amplitudes are all proportional to $1/f^{3}$,
where $f$ is the light pseudoscalar meson decay constant
in the chiral limit.  At NLO, we choose to express $1/f^{3}$ in
terms of $1/(f^{2}_{\pi} f_{K})$.  This factor fully absorbs
the dependence upon the Gasser-Leutwyler LECs $L_{4}$ and $L_{5}$
introduced via wavefunction renormalisation.

\smallskip

In Ref. \cite{Lin:2002nq}, the one-loop diagrams have been 
calculated for arbitrary external 
meson four-momenta in both $\chi$PT and q$\chi$PT.  
The results are
lengthy and are presented on a web site \cite{website}.
In Fig. 2, we show an example of these results for 
$\la\pi^{+}\pi^{0}|\op_{7,8}|K^{+}\ra$.  These plots are
the ratios between the one-loop corrections and
the lowest-order matrix elements with both
final-state pions at rest.  Fig. 2a is the result in $\chi$PT and
Fig. 2b is that in q$\chi$PT.   This figure suggests that in
a quenched numerical calculation of these matrix elements, 
it is not appropriate in general to perform chiral extrapolation using 
unquenched $\chi$PT results \cite{panel:lattice02}.  
This is confirmed by numerical data \cite{Papinutto:latt02}.
\begin{center}
\begin{figure}
\vspace{0.0cm}
\epsfxsize=7cm
\epsffile{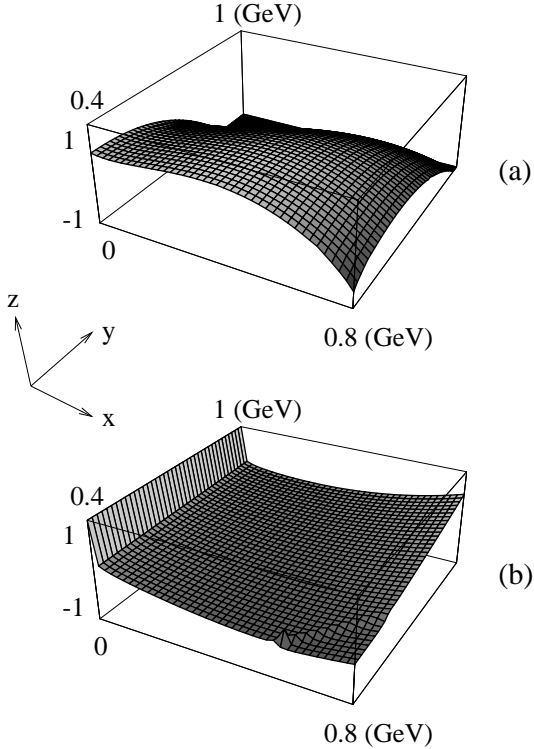}
\vspace{0.0cm}
\label{fig:logslambda07}
\caption{{\sl Ratio 
between the one-loop correction, at the renormalisation
scale 0.7 GeV, and the lowest-order result for
$\la\pi^{+}\pi^{0}|\op_{7,8}|K^{+}\ra$ in (a) $\chi$PT and (b) 
q$\chi$PT.
The x axis is $M_{\pi}$ and the y axis is $M_{K}$.  In (b), the 
coupling accompanying the
kinetic term of the $\eta^{\prime}$ propagator is set to zero, and the 
$\eta^{\prime}$ mass is taken to be $M_{0}=0.5$ GeV.  The one-loop
results are not very sensitive to these parameters.
The singular 
behaviour along the line $M_{\pi}=\sqrt{2}M_{K}$ in (b) is 
due to the fact
that when performing the q$\chi$PT calculation, we use a basis in 
which the pseudo-Goldstone states are $\bar{q}q^{\prime}$ mesons, 
where $q$ and $q^{\prime}$ are $u, d$ and $s$, and the $\bar{s}s$ meson
becomes tachyonic when $M_{\pi} > \sqrt{2}M_{K}$.
}}
\end{figure}
\end{center}

\section{CONCLUSIONS}
We have made theoretical progress towards the calculation of 
$K\rightarrow\pi\pi$ decay amplitudes via the combination of lattice
QCD and $\chi$PT. We find it
feasible to determine all the LECs for constructing
$_{I=2}\la\pi\pi|\op^{\Delta S = 1}|K\ra$ at NLO in the chiral expansion.  
A quenched numerical
study is in progress \cite{Papinutto:latt02}.  As for the 
$\Delta I =1/2$ channel, we find the situation to be considerably 
more complicated
\cite{Villadoro:latt02}.

\end{document}